\documentclass[sigconf]{acmart}
\AtBeginDocument{%
  }

\usepackage{algorithm}
\usepackage{algpseudocode}
\usepackage{amsmath}
\usepackage{graphicx}
\usepackage{geometry}
\usepackage{xcolor}
\usepackage{colortbl}
\begin{document}

\title{AgriVariant: Variant Effect Prediction using DeepChem-Variant 
for Precision Breeding in Rice}

\author{Ankita Vaishnobi Bisoi}
\email{f20212306@goa.bits-pilani.ac.in}
\affiliation{%
  \institution{BITS Pilani Goa Campus, India}
  \state{Goa}
  \country{India}
}

\author{Bharath Ramsundar}
\email{bharath@deepforestsci.com}
\affiliation{%
  \institution{Deep Forest Sciences}
  \state{California}
  \country{USA}}

\renewcommand{\shortauthors}{Bisoi et al.}

\begin{abstract}
Predicting functional consequences of genetic variants in crop genes remains a critical bottleneck for precision breeding programs. We present AgriVariant, an end-to-end pipeline for variant-effect prediction in rice (Oryza sativa) that addresses the lack of crop-specific variant-interpretation tools and can be extended to any crop species with available reference genomes and gene annotations. Our approach integrates deep learning-based variant calling (DeepChem-Variant \cite{DeepChemVariant}) with custom plant genomics annotation using RAP-DB\cite{Sakai2013RAPDB}\cite{Kawahara2013NipponbareRAPDB} gene models and database-independent deleteriousness scoring that combines the Grantham\cite{Grantham1974} distance and the BLOSUM62 \cite{Henikoff1992BLOSUM} substitution matrix. We validate the pipeline through targeted mutations in stress-response genes (OsDREB2a, OsDREB1F, SKC1), demonstrating correct classification of stop-gained, missense, and synonymous variants with appropriate HIGH / MODERATE / LOW impact assignments. An exhaustive mutagenesis study of OsMT-3a analyzed all 1,509 possible single-nucleotide variants in 10 days, identifying 353 high-impact, 447 medium-impact, and 709 low-impact variants —an analysis that would have required 2-4 years\cite{collard2017revisiting} using traditional wet-lab approaches. This computational framework enables breeders to prioritize variants for experimental validation across diverse crop species, reducing screening costs and accelerating development of climate-resilient crop varieties.
\end{abstract}



\maketitle

\section{Introduction}

Climate change threatens global food security through increased frequency and severity of environmental stresses. Rice (Oryza sativa), a staple crop feeding over half the world's population, faces particular vulnerability to drought and salinity stress. Genetic variants in stress-response genes dictate crop durability, yet identifying which variants functionally impact protein activity remains a critical challenge despite vast available genomic data.

We address this challenge by developing AgriVariant, a variant effect prediction pipeline that integrates deep learning-based variant calling with plant-specific functional annotation and quantitative deleteriousness scoring. Building upon our previous work on DeepChem-Variant \cite{DeepChemVariant}, a modular implementation of DeepVariant \cite{poplin2017practical}\cite{poplin2018universal} within the DeepChem \cite{Ramsundar-et-al-2019} framework, we extend variant calling capabilities to downstream functional interpretation specifically for crop genomes. DeepChem-Variant leverages convolutional neural networks \cite{krizhevsky2012imagenet} to reframe variant calling as image classification, generating multi-channel pileup images from sequencing reads and processing them through Inception V3 \cite{szegedy2015rethinkinginceptionarchitecturecomputer} or MobileNetV2 \cite{mobilenetv2} architectures, achieving 95.3\% accuracy on rice genomic variants.

DeepChem, the underlying framework, is an open-source Python library designed for scientific machine learning that has established itself as a versatile platform for applications ranging from the MoleculeNet benchmark suite \cite{wu2018moleculenet} to protein-ligand interaction modeling \cite{gomes2017atomic} and generative modeling of molecules \cite{frey2022fastflows}. This ecosystem enables AgriVariant to leverage mature molecular machine learning infrastructure while providing modular, extensible components specifically tailored for crop genomics applications.

The primary bottleneck in crop breeding lies not in variant detection, but in variant interpretation. We target stress-response genes including the DREB transcription factor family (OsDREB2a, OsDREB1F) \cite{osdreb2a} \cite{osdreb1f}, ion homeostasis genes (SKC1) \cite{skc1}, and metallothionein genes (OsMT-3a) \cite{osmt3a}, which collectively regulate drought tolerance, salinity stress, and heavy metal detoxification in rice. However, existing variant effect prediction tools (SIFT\cite{Sim2012SIFT}, PolyPhen-2\cite{polyphen2}, CADD\cite{Kircher2014CADD}) rely heavily on human-specific databases and evolutionary conservation patterns, leaving plant genomics without equivalent prediction frameworks. Existing plant annotation tools like SnpEff\cite{snpeff} provide basic functional classification but offer limited species coverage and lack crop-specific databases for many reference genomes.

We overcome this limitation by developing a custom functional annotation pipeline that maps variants to coding sequences using  Rice Annotation Project Database (RAP-DB)\cite{Sakai2013RAPDB}\cite{Kawahara2013NipponbareRAPDB} gene models and classifies functional effects (stop-gained, missense, synonymous) based on affected codons and amino acid translations. We implement a database-independent deleteriousness scoring strategy that merges Grantham\cite{Grantham1974} distance matrices (which quantify amino acid divergence based on composition, polarity, and volume) with BLOSUM62 \cite{Henikoff1992BLOSUM} substitution matrices (reflect evolutionary tolerance to amino acid changes). This combined scoring framework enables quantitative variant severity assessment without requiring organism-specific training databases, proving particularly valuable for understudied crops where extensive variant annotations are unavailable.

\section{Methods}

\begin{figure}[H]
    \centering
    \includegraphics[width=1\columnwidth]{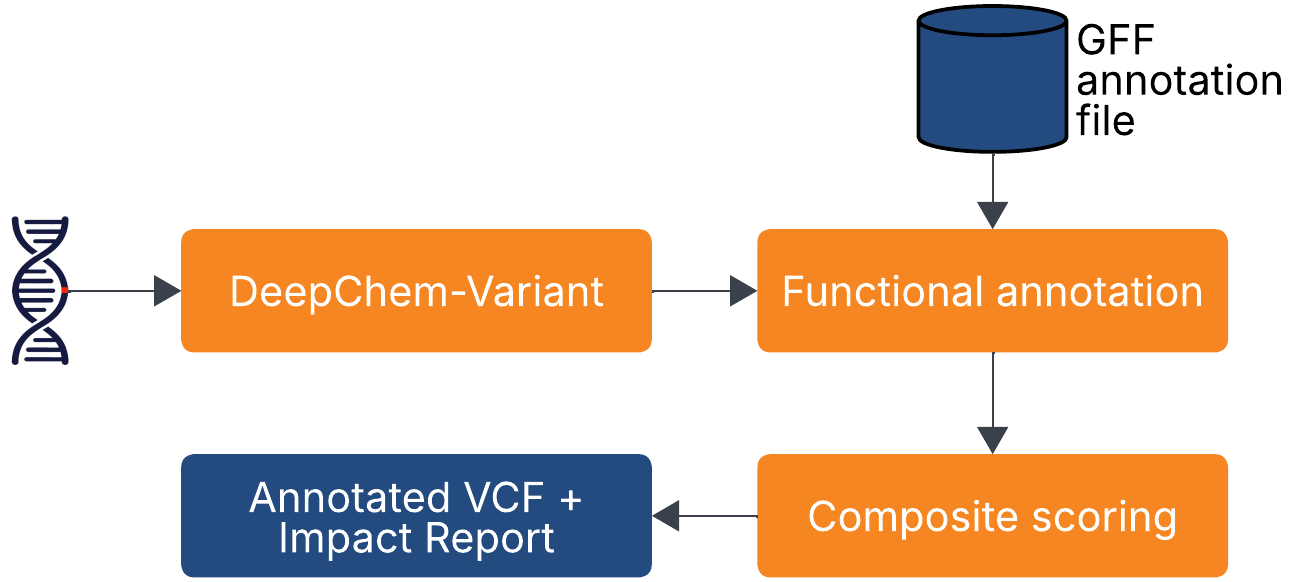}
    \caption{AgriVariant pipeline integrating deep learning-based variant calling, plant-specific functional annotation, and quantitative deleteriousness scoring.}
    \label{fig:pipeline-overview}
\end{figure}

The complete pipeline for AgriVariant (Figure \ref{fig:pipeline-overview}) combines three complementary methods for variant interpretation in rice genomes. Sequencing data undergoes variant calling via DeepChem-Variant, followed by functional annotation using custom scripts that map mutations to coding sequences and classify effects based on RAP-DB gene models. Detected variants are then scored for deleteriousness using a composite metric derived from Grantham distance and BLOSUM62 substitution matrices. The output consists of annotated Variant Call Format (VCF) files containing variant positions, functional classifications, and quantitative severity scores. This modular architecture enables independent updating of individual components while maintaining compatibility with standard genomic file formats.

\vspace{-0.9cm}

\subsection{Target Gene Selection and Biological Context}

We selected target genes based on documented roles in abiotic stress tolerance and ion homeostasis. The DREB (Dehydration-Responsive Element-Binding) gene family belongs to the AP2/ERF superfamily of plant transcription factors. These genes regulate responses to abiotic stresses(drought, cold, and salt) by binding to DRE/CRT cis-elements in DNA, thereby activating downstream stress-responsive genes and improving plant tolerance. 

\begin{figure}[H]
    \centering
    \includegraphics[width=0.85\columnwidth]{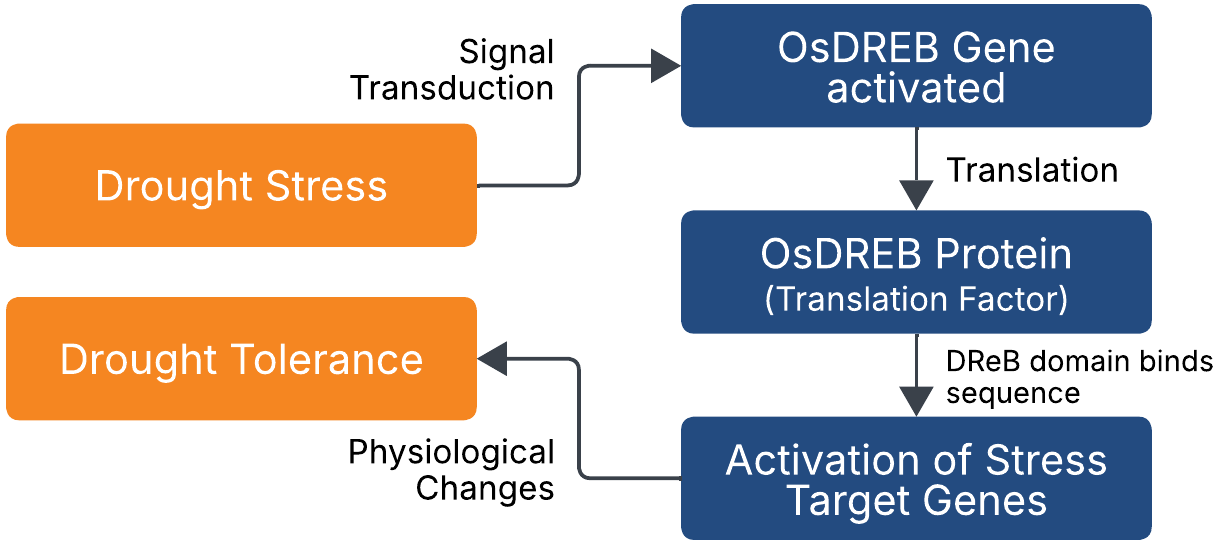}
    \caption{OsDREB-mediated drought stress response pathway includes stress-induced gene activation, transcription factor production, downstream target gene regulation, and resulting drought tolerance.}
    \label{fig:OsDREB}
\end{figure}

Two DREB family members, OsDREB2a \cite{osdreb2a} and OsDREB1F \cite{osdreb1f}, were selected as primary targets for loss-of-function variant analysis. When a plant experiences stress, expression of OsDREB2a and OsDREB1F is induced (Figure \ref{fig:OsDREB}). The resulting proteins act as transcriptional activators, switching on expression of target genes. This activation leads to accumulation of osmolytes such as soluble sugars and free proline, which help the plant adjust osmotic potential, maintain cell structure, and protect cellular components under stress.

\begin{figure}[H]
    \centering
    \includegraphics[width=0.85\columnwidth]{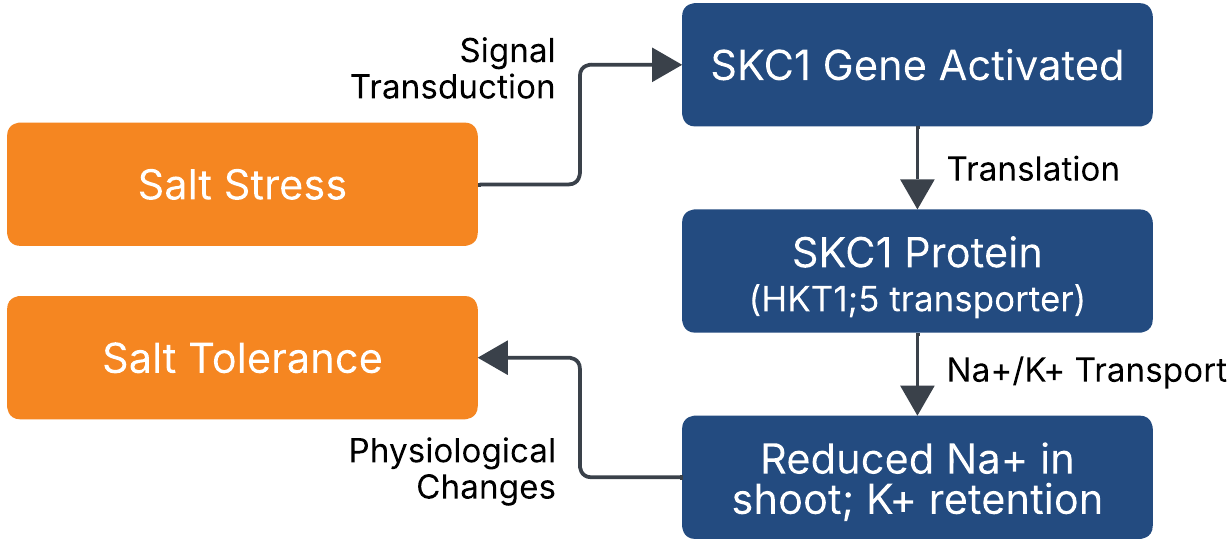}
    \caption{SKC1-mediated salt stress response pathway. Salt stress activates 
    SKC1 gene expression, producing HKT1;5 transporter protein that regulates 
    Na+/K+ homeostasis by reducing sodium accumulation in shoots while 
    maintaining potassium levels, resulting in salt tolerance.}
    \label{fig:SKC}
\end{figure}

\begin{figure}[H]
    \centering
    \includegraphics[width=0.85\columnwidth]{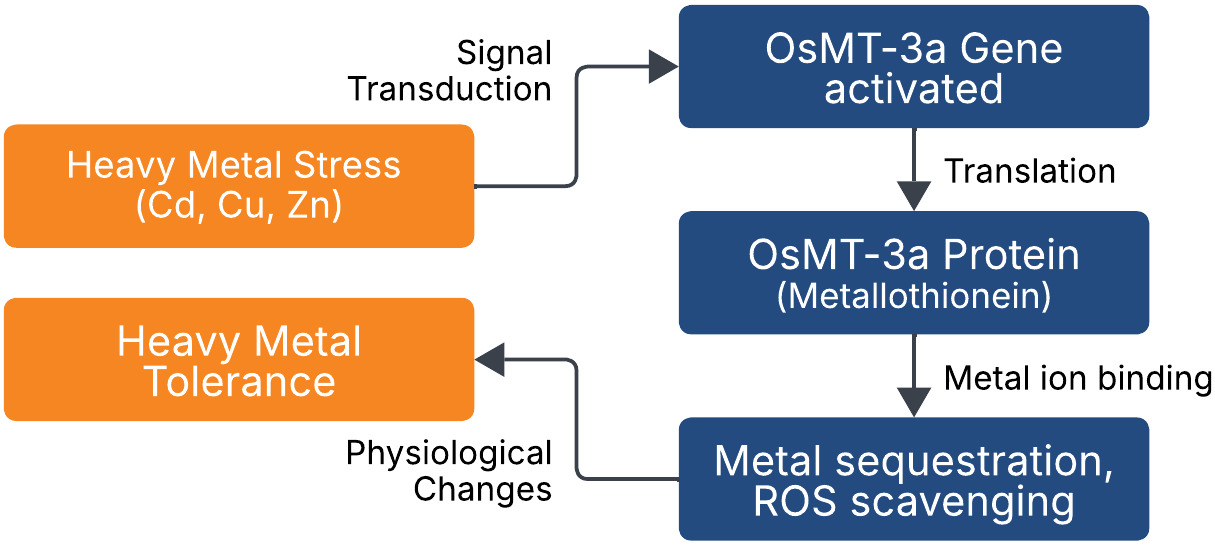}
    \caption{OsMT-3a-mediated heavy metal stress response pathway. Heavy metal 
    exposure (Cd, Cu, Zn) induces OsMT-3a expression, producing metallothionein 
    protein that binds and sequesters toxic metal ions while scavenging reactive 
    oxygen species, conferring heavy metal tolerance.}
    \label{fig:osmt3a}
\end{figure}

Beyond transcriptional regulation, ion homeostasis represents another critical mechanism for stress tolerance. SKC1 (Shoot K+ Concentration 1), also known as OsHKT1;5, is a high-affinity K+ transporter that regulates sodium-potassium balance in rice tissues during salt stress \cite{skc1}. Genetic variants in SKC1 influence salt tolerance (Figure \ref{fig:SKC}) across rice cultivars, making it an ideal candidate for neutral variant analysis. Additionally, we examined OsMT-3a \cite{osmt3a}, a metallothionein gene involved in heavy metal detoxification and oxidative stress responses (Figure \ref{fig:osmt3a}), to demonstrate AgriVariant's applicability across diverse stress-response mechanisms.

\subsection{Computational Mutation Framework}

\textbf{Rationale for \textit{in silico} Approach:} Traditional functional validation of genetic variants requires laboratory mutagenesis using CRISPR-Cas9 gene editing, followed by phenotypic screening across multiple growing seasons—a process requiring 2-4 years and significant resources. For breeding programs evaluating thousands of candidate variants across multiple genes, this approach becomes expensive and time-consuming. AgriVariant's computational mutation framework enables rapid \textit{in silico} screening of variant effects, allowing breeders to prioritize the most promising candidates for subsequent wet-lab validation.

We implemented a synthetic variant generation pipeline that mimics the CRISPR mutagenesis workflow computationally (Algorithm \ref{alg:mutation-framework}). This approach creates artificial genomes containing targeted mutations, generates synthetic sequencing reads from these modified genomes, and processes them through DeepChem-Variant to validate detection accuracy and annotation correctness.

\begin{algorithm}[H]
\caption{Synthetic Variant Generation and Validation Framework}
\label{alg:mutation-framework}
\begin{algorithmic}[1]
\Require Reference genome (FASTA), target gene coordinates, desired mutation
\Ensure Validated variant in synthetic dataset

\State \textbf{// Phase 1: Mutation Design}
\State Extract CDS coordinates from GFF annotation for target gene
\State Identify target codon position and genomic coordinate
\State Design mutation: $(chrom, pos, ref\_allele, alt\_allele)$
\State Verify mutation type: stop-gained, missense, or synonymous

\State
\State \textbf{// Phase 2: Synthetic Genome Generation}
\State Create VCF file with mutation: $\{chrom, pos, ref, alt, GT=1/1\}$
\State Compress and index VCF: \texttt{bgzip}, \texttt{tabix}
\State Generate mutant genome: \texttt{bcftools consensus} $ref.fa$ $mutation.vcf.gz$
\State Index mutant genome: \texttt{samtools faidx}

\State
\State \textbf{// Phase 3: Synthetic Read Generation}
\State Extract target region: $[gene\_start - 5kb, gene\_end + 5kb]$
\State Generate paired-end reads: \texttt{wgsim -N 1000 -1 150 -2 150}

\State
\State \textbf{// Phase 4: Variant Calling Validation}
\State Align synthetic reads to original reference: \texttt{bwa mem}
\State Sort and index BAM file: \texttt{samtools sort}, \texttt{samtools index}
\State Verify mutation presence: \texttt{samtools mpileup} at target position
\State Run DeepChem-Variant pipeline (Algorithm \ref{alg:deepchem-variant})

\State
\State \textbf{// Phase 5: Validation}
\If{called variant matches designed mutation}
    \State \textbf{Success:} Mutation correctly detected
\Else
    \State \textbf{Failure:} Investigate detection/annotation errors
\EndIf

\State \Return Annotated variant with functional classification
\end{algorithmic}
\end{algorithm}

AgriVariant serves as a proof-of-concept for computational variant prioritization. In production breeding workflows, researchers would use these predictions to rank thousands of naturally occurring or CRISPR-induced variants, selecting only the highest-impact candidates for costly field trials. 

\subsection{Variant Calling with Deep-Chem-Variant}

\begin{figure}[H]
    \centering
    \includegraphics[width=0.8\columnwidth]{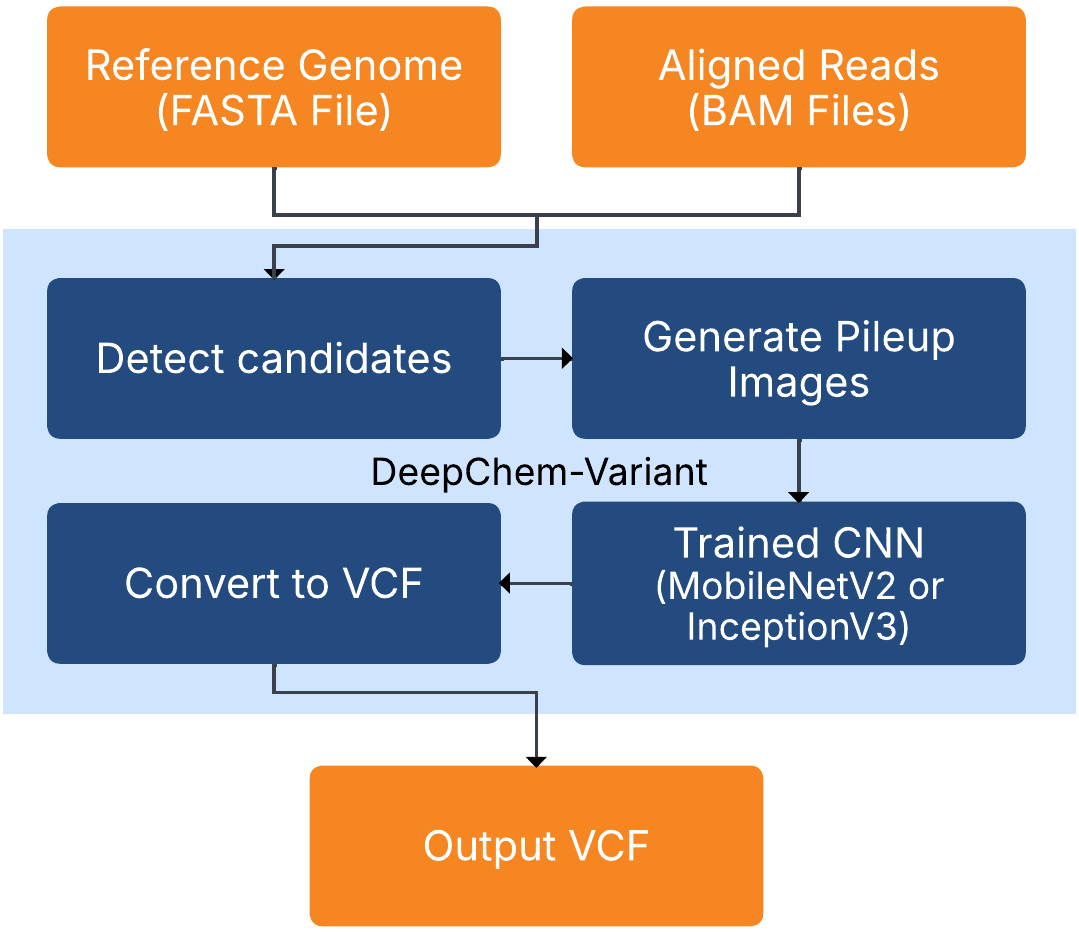}
    \caption{DeepChem-Variant workflow: Reference genome and aligned reads are processed through candidate detection, pileup image generation (6-channel tensors), CNN classification (MobileNetV2 or InceptionV3), and VCF conversion to produce annotated variant calls.}
    \label{fig:deepchem-variant}
\end{figure}

DeepChem-Variant performed alignment-based variant calling between reference and simulated genomes. Unlike traditional statistical variant callers (GATK\cite{gatk}, SAMtools\cite{samtools}) that rely on hand-crafted probabilistic models, DeepChem-Variant leverages convolutional neural networks to learn complex inter-read dependencies directly from data (Figure \ref{fig:deepchem-variant}).

DeepChem-Variant converts aligned sequencing reads into multi-channel pileup images (6 channels encoding base identity, base quality, mapping quality, strand orientation, variant support, and reference match indicators across 221-bp windows with 100-read depth). These images are processed through a MobileNetV2 or InceptionV3 CNN that outputs probabilistic genotype classifications: homozygous reference, heterozygous, or homozygous alternate (Algorithm \ref{alg:deepchem-variant}). This image-based representation enables the model to capture spatial patterns in read alignments that traditional methods cannot detect.

\vspace{-0.4cm}

AgriVariant is built within the DeepChem framework and is fully open source, with the entire codebase written in Python. This design enables species specific optimization that is not possible with closed source or human centric variant callers. We trained the model on Indica rice genomic data, and the same architecture allows researchers to retrain it on their own crop varieties, adapt it to new sequencing platforms, and integrate variant calling directly with downstream molecular machine learning workflows. These capabilities are especially important for resource limited breeding programs working with understudied crops.

\begin{algorithm}[H]
\caption{DeepChem-Variant}
\label{alg:deepchem-variant}
\begin{algorithmic}[1]
\Require Aligned reads (BAM $B$), Reference genome (FASTA $F$)
\Ensure Variants with genotype probabilities (VCF)

\State $\mathcal{C} \gets \text{CandidateFeaturizer}(B, F)$ 
\Comment{Realign, detect candidates}
\State $\mathcal{I} \gets \text{PileupFeaturizer}(\mathcal{C})$ 
\Comment{Generate $6 \times 100 \times 221$ tensors}
\State $\mathbf{P} \gets \text{CNN}_{\text{MobileNetV2}}(\mathcal{I})$ 
\Comment{$[P(\text{0/0}), P(\text{0/1}), P(\text{1/1})]$}
\State Filter variants where $\arg\max(\mathbf{P}) \neq \text{hom-ref}$
\State Compute quality: $Q = -10\log_{10}(1 - \max(\mathbf{P}))$
\State \Return VCF with genotypes and quality scores
\end{algorithmic}
\end{algorithm}

Variant calling generated Variant Call Format (VCF) files containing chromosome positions, reference and alternate alleles, quality scores, and filter status for each detected variant. We validated detection accuracy by confirming that simulated variants appeared in output VCFs with correct genomic coordinates and allelic states. 

\subsection{Functional Annotation}

Absence of crop-specific databases in standard annotation tools (SnpEff) necessitated development of custom annotation infrastructure. We obtained gene structure information from RAP-DB  in General Feature Format (GFF) files containing exon coordinates, coding sequence (CDS) boundaries, and transcript identifiers for all annotated rice genes.

\begin{figure}[H]
    \centering
    \includegraphics[width=0.9\columnwidth]{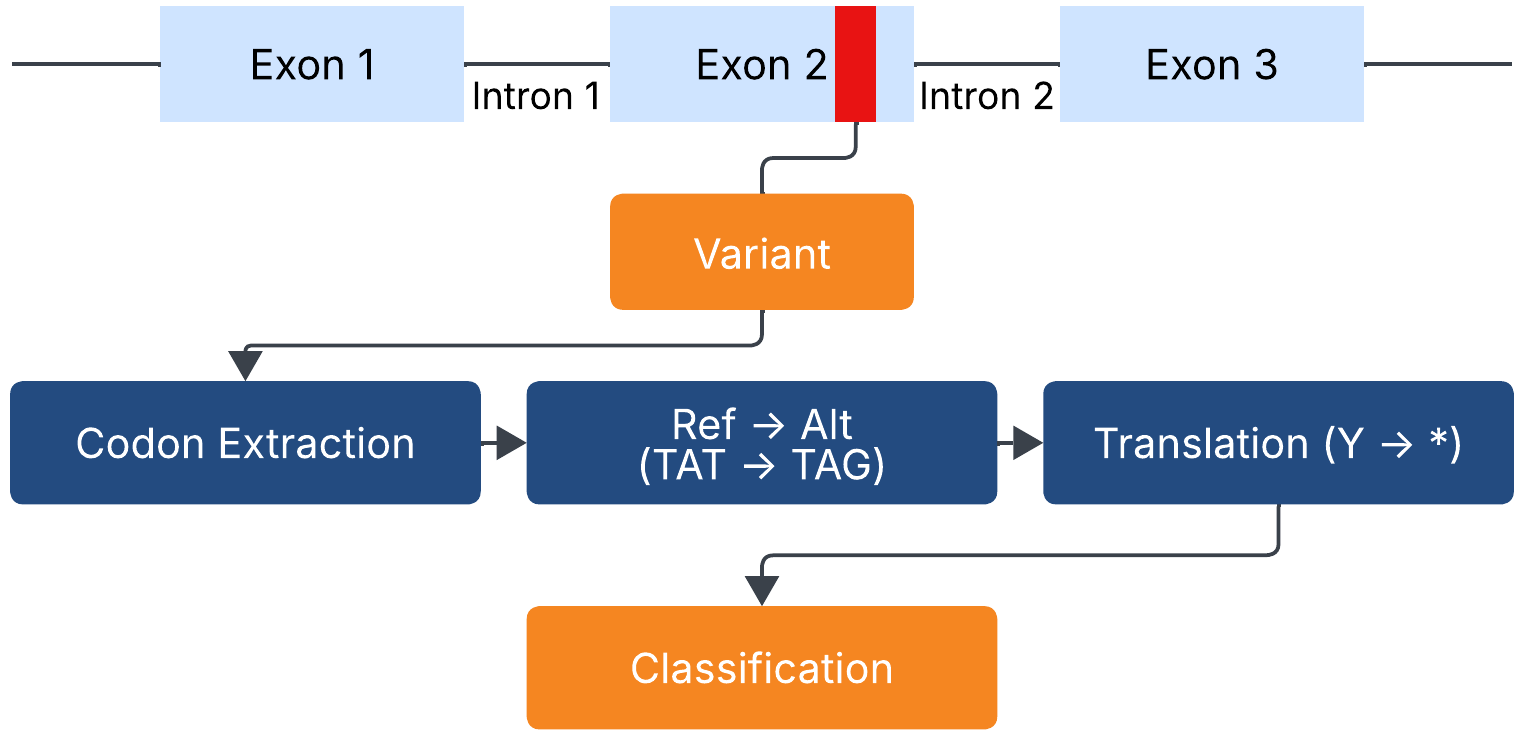}
    \caption{Functional annotation workflow mapping variants from genomic position through codon extraction and translation to amino acid change}
    \label{fig:functional_annotation}
\end{figure}

Our annotation pipeline (Figure \ref{fig:functional_annotation}) parsed GFF files to extract CDS coordinates for target genes and mapped variant positions from VCF files to corresponding exonic regions. For each detected variant, the algorithm identified overlapping CDS features based on chromosome and genomic coordinates, calculated relative position within the CDS while accounting for strand orientation, and determined the affected codon. The reference codon sequence was retrieved from the genome FASTA file, the variant nucleotide substitution was applied to generate the alternate codon, and both codons were translated using the standard genetic code. Variants were classified as missense, nonsense, or silent based on resulting amino acid changes.

The codon extraction process accounted for multi-exon genes by retrieving nucleotides from adjacent exons to assemble complete codons for variants near exon boundaries. Strand orientation determined whether codons were read 5' to 3' or reverse complemented before translation. Functional effect classification followed standard nomenclature: stop-gained variants introduced premature stop codons, missense variants changed amino acid identity, and silent (synonymous) variants maintained the same amino acid despite nucleotide changes. Impact severity was assigned using established guidelines: HIGH for stop-gained and frameshift variants, MODERATE for missense variants, and LOW for synonymous variants.

\subsection{Database-Independent Deleteriousness Scoring}

Most widely used deleteriousness prediction tools (SIFT, PolyPhen-2, CADD) depend heavily on human-specific reference databases and evolutionary conservation patterns derived from animal genomes, limiting their applicability to plant genomes. To enable quantitative assessment of variant severity in rice genes, we implemented a database-independent scoring strategy based on amino acid physicochemical properties and evolutionary substitution likelihoods, making this an approach that generalizes to any organism without requiring pre-trained species-specific models.

\begin{figure}[H]
    \centering
    \includegraphics[width=0.95\columnwidth]{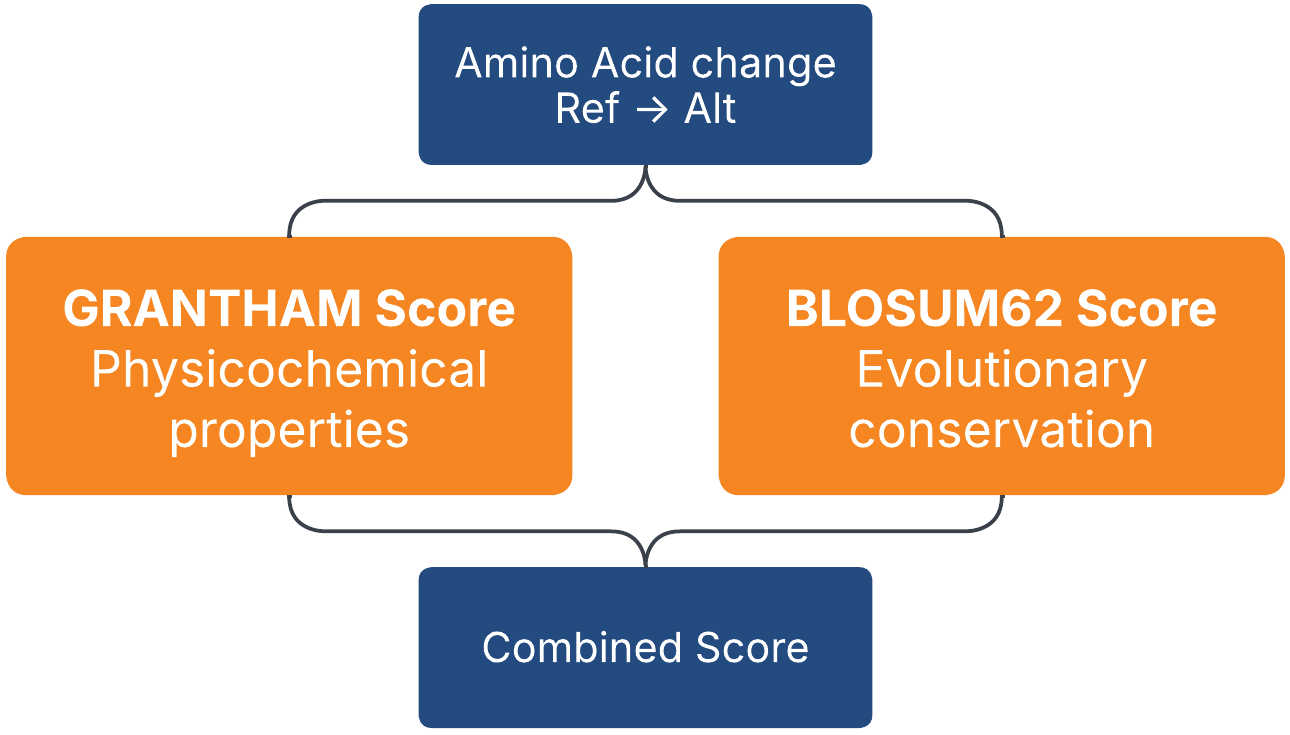}
    \caption{Composite Scoring combining Grantham Score and BLOSUM62 Score}
    \label{fig:scoring}
\end{figure}

We combined two complementary scoring metrics (Figure \ref{fig:scoring}). The Grantham distance matrix quantifies amino acid dissimilarity based on composition (atomic weight ratios of non-carbon elements), polarity (hydrophobicity measurements), and molecular volume. Grantham scores range from 5 (conservative substitutions) to 215 (radical substitutions). Reference and alternate amino acids from annotation output were used to query the Grantham matrix and retrieve distance scores.

BLOSUM62 (BLOcks SUbstitution Matrix 62) derives from observed substitution frequencies in protein alignments sharing at least 62\% sequence identity. Positive scores indicate substitutions occurring more frequently than expected by chance (conservative), whereas negative scores correspond to substitutions less favored in nature (radical). Stop-gained mutations are scored 1 by AgriVariant as they introduce premature stop codons with HIGH impace severity. This metric estimates evolutionary tolerance of specific amino acid changes.

\vspace{-.2cm}

To enable integration of both metrics, Grantham scores were normalized to 0–1 range by dividing by maximum observed value (215), while BLOSUM62 scores were normalized by shifting by minimum matrix value and dividing by overall score range. The resulting composite score provides a quantitative measure of variant severity independent of categorical impact classifications:

\[
\text{Composite Score} = \frac{1}{2}\left(\frac{\text{Grantham}}{215} + \frac{\text{BLOSUM62}_{\text{min}} - \text{BLOSUM62}}{\text{BLOSUM62}_{\text{range}}}\right)
\]

By using this framework, rather than training organism-specific machine learning models (which require extensive labeled data), we leverage universal biochemical principles encoded in substitution matrices. This approach provides immediate generalizability to understudied crops without requiring training data.



\section{Case Studies}

To validate AgriVariant's performance, we obtained BAM files for three rice mutant lines the MiRiQ Database \cite{Kubo2024MiRiQ} and analyzed them. These variants span the functional impact spectrum—MODERATE impact missense substitutions (OsDREB2a, SKC1) and a LOW impact synonymous control (OsDREB1F), thus demonstrating the pipeline's ability to correctly classify variant effects across diverse stress-response genes and severity levels.

\subsection{OsDREB2a Missense Mutation}

The variant Os01t0165000-01:c.794G$>$A (p.G265E) introduced a guanine-to-adenine substitution at nucleotide position 794, converting a glycine codon to glutamate at amino acid position 265 (Table \ref{tab:dreb2a_variant}). AgriVariant successfully detected this substitution, reporting it in the VCF at position chr01:3358191 with reference allele G and alternate allele A.

 The substitution replaces glycine, a small non-polar amino acid that confers conformational flexibility, with glutamate, a larger negatively charged residue. This physicochemical change scored moderately severe on both substitution matrices: Grantham distance of 98 (on a 0-215 scale where higher values indicate more radical changes) and BLOSUM62 score of -2 (negative values indicate substitutions disfavored by evolution). The combined deleteriousness score of 0.137 reflected the intermediate severity of this substitution.

AgriVariant correctly classified this variant as missense with MODERATE impact severity. The glycine-to-glutamate substitution at position 265 introduces both steric bulk and electrostatic charge in a region that may affect protein stability or regulatory interactions. The missense variant likely reduces but does not eliminate OsDREB2a activity, potentially resulting in intermediate drought tolerance phenotypes. Plants carrying this variant would likely exhibit reduced drought tolerance compared to wild-type but retain partial stress response capacity, making it a candidate for functional validation studies.

\begin{table}[!t]
\centering
\caption{AgriVariant validation of a missense OsDREB2a variant. Simulated gene context (top) and automated functional annotation with MODERATE-impact classification (bottom). Green highlights denote correct predictions.}
\label{tab:dreb2a_variant}
\begin{tabular}{@{}ll@{}}
\toprule
\multicolumn{2}{c}{\textbf{Background Context (Known from Simulation)}} \\
\midrule
Gene ID & Os01g0165000 \\
Gene Name & DREB2A \\
Chromosome:Position & \cellcolor{green!50} chr01:3358191 \\
Ref $\rightarrow$ Alt Nucleotide & \cellcolor{green!50} G $\rightarrow$ A \\
\midrule
\multicolumn{2}{c}{\textbf{AgriVariant Predictions (Validated Outputs)}} \\
\midrule
Ref $\rightarrow$ Alt Codon  & \cellcolor{green!50} G\fbox{G}G $\rightarrow$ G\fbox{A}G \\
Amino Acid Change & \cellcolor{green!50} G $\rightarrow$ E (Glycine $\rightarrow$ Glutamate) \\
Grantham Score & \cellcolor{green!50} 98 \\
BLOSUM62 Score & \cellcolor{green!50} -2 \\
Combined Score & \cellcolor{green!50} 0.137 \\
Functional Effect & \cellcolor{green!50} Missense \\
Impact Severity & \cellcolor{green!50} MODERATE \\
\bottomrule
\end{tabular}
\end{table}

\subsection{OsDREB1F Synonymous Mutation}

The variant Os01g0968800-01:c.165G$>$A (p.G55G) replaces guanine with adenine at nucleotide position 165, maintaining glycine at amino acid position 55 (Table \ref{tab:dreb1f_variant}). AgriVariant successfully detected this substitution, reporting it in the VCF at position chr01:42727659 with reference allele G and alternate allele A.

Annotation analysis identified the variant within the OsDREB1F coding sequence. The substitution altered the third position of the glycine codon, changing GGG to GGA. Both of these encode the same amino acid due to the genetic code's degeneracy. This represents a synonymous or silent mutation where the nucleotide sequence changes but the protein sequence remains identical.

AgriVariant correctly classified this variant as synonymous with LOW impact severity. The Grantham score of 0 reflects no amino acid change (glycine to glycine), while the BLOSUM62 score of 7 (the maximum value, indicating identical amino acids) confirms the neutral nature of this substitution. 

This control variant validates the pipeline's ability to distinguish functionally neutral changes from deleterious mutations. Plants carrying this variant would exhibit normal OsDREB1F function, as the protein sequence remains unchanged despite the genomic alteration.

\begin{table}[!t]
\centering
\caption{AgriVariant validation of a synonymous OsDREB1F variant. Simulated gene context (top) and automated functional annotation with LOW-impact classification (bottom). Green highlights denote correct predictions.}
\label{tab:dreb1f_variant}

\begin{tabular}{@{}ll@{}}
\toprule
\multicolumn{2}{c}{\textbf{Background Context (Known from Simulation)}} \\
\midrule
Gene ID & Os01g0968800 \\
Gene Name & DREB1F \\
Chromosome:Position & \cellcolor{green!50} chr01:42727659 \\
Ref $\rightarrow$ Alt Nucleotide & \cellcolor{green!50} G $\rightarrow$ A \\
\midrule
\multicolumn{2}{c}{\textbf{AgriVariant Predictions (Validated Outputs)}} \\
\midrule
Ref $\rightarrow$ Alt Codon & \cellcolor{green!50} GG\fbox{G} $\rightarrow$ GG\fbox{A} \\
Amino Acid Change & \cellcolor{green!50} G $\rightarrow$ G (Glycine $\rightarrow$ Glycine) \\
Grantham Score & \cellcolor{green!50} 0 \\
BLOSUM62 Score & \cellcolor{green!50} 7 \\
Combined Score & \cellcolor{green!50} -0.50 \\
Functional Effect & \cellcolor{green!50} Synonymous \\
Impact Severity & \cellcolor{green!50} LOW \\
\bottomrule
\end{tabular}
\end{table}

\subsection{SKC1 Missense Mutation}

The SKC1 variant Os01t0307500-01:c.1075T$>$A (p.C359S) replaced thymine with adenine at nucleotide position 1075, converting cysteine at position 359 to serine (Table \ref{tab:SKC1_variant}). AgriVariant successfully detected this substitution, reporting it in the VCF at position chr01:11462201 with reference allele T and alternate allele A.

Annotation analysis identified the variant within the SKC1 coding sequence at amino acid position 359. The substitution replaces cysteine, a sulfur-containing amino acid capable of forming disulfide bonds, with serine, a polar hydroxyl-containing residue. While both are small polar amino acids, the loss of cysteine's sulfur group eliminates potential disulfide bond formation, which could affect protein structure if position 359 participates in structural stabilization. The Grantham score of 112 indicates moderate physicochemical divergence, while the BLOSUM62 score of -1 suggests this substitution is slightly disfavored in evolution. The combined deleteriousness score of 0.124 reflected intermediate severity.

AgriVariant correctly classified this variant as missense with MODERATE impact severity. The cysteine-to-serine substitution at position 359 represents a moderately conservative change (both residues are small and polar) but the functional consequences depend on whether this cysteine participates in disulfide bonding or active site chemistry. In the context of the HKT1;5 sodium transporter, this position may influence ion selectivity or transport kinetics, potentially affecting salt tolerance phenotypes.

\begin{table}[!t]
\centering
\caption{AgriVariant validation of a missense SKC1 variant. Simulated gene context (top) and automated functional annotation with MODERATE-impact classification (bottom). Green highlights denote correct predictions.}
\label{tab:SKC1_variant}
\begin{tabular}{@{}ll@{}}
\toprule
\multicolumn{2}{c}{\textbf{Background Context (Known from Simulation)}} \\
\midrule
Gene ID & Os01g0307500 \\
Gene Name & SKC1 \\
Chromosome:Position & \cellcolor{green!50} chr01:11462201 \\
Ref $\rightarrow$ Alt Nucleotide & \cellcolor{green!50} T $\rightarrow$ A \\
\midrule
\multicolumn{2}{c}{\textbf{AgriVariant Predictions (Validated Outputs)}} \\
\midrule
Ref $\rightarrow$ Alt Codon & \cellcolor{green!50} \fbox{T}GT $\rightarrow$ \fbox{A}GT \\
Amino Acid Change & \cellcolor{green!50} C $\rightarrow$ S (Cysteine $\rightarrow$ Serine) \\
Grantham Score & \cellcolor{green!50} 112 \\
BLOSUM62 Score & \cellcolor{green!50} -1 \\
Combined Score & \cellcolor{green!50} 0.124 \\
Functional Effect & \cellcolor{green!50} Missense \\
Impact Severity & \cellcolor{green!50} MODERATE \\
\bottomrule
\end{tabular}
\end{table}

\section{Exhaustive Mutation Analysis of OsMT-3a}

\subsection{Rationale and Gene Selection}

OsMT-3a encodes a type 3 metallothionein involved in heavy metal detoxification and oxidative stress response in rice \cite{osmt3a}. Metallothioneins are cysteine-rich proteins that bind and sequester toxic metal ions (cadmium, copper, zinc), protecting cells from oxidative damage. The gene's 183-bp coding sequence produces a compact 61-amino acid protein, making it an ideal candidate for exhaustive computational mutagenesis.

Traditional breeding approaches to assess all possible single-nucleotide variants in OsMT-3a would require generating 1,509 individual mutant lines (accounting for all substitutions, insertions, and deletions), followed by 8-10 growing seasons for phenotypic evaluation. The selection and screening would span 2-4 years and require extensive greenhouse facilities. Our computational pipeline completed this analysis in approximately 10 days on standard hardware, demonstrating its utility for rapid variant prioritization in breeding programs.

We systematically generated all possible single-nucleotide substitutions, insertions, and deletions across the 183-bp OsMT-3a coding sequence. For each mutation, a synthetic variant was introduced into the reference genome using \texttt{bcftools consensus}, followed by generation of synthetic sequencing reads using \texttt{wgsim} with 1000× coverage. Variants were detected using DeepChem-Variant (Algorithm \ref{alg:deepchem-variant}), annotated for functional effects using the custom plant genomics pipeline, and scored for deleteriousness via the Grantham-BLOSUM62 composite metric.

Variants were classified into three deleteriousness categories based on composite scores: High (score > 0.3), Medium (0 < score <= 0.3), and Low (score <= 0). Functional effects were categorized as missense (amino acid substitution), synonymous (silent mutation), nonsense (premature stop codon), or stop loss (stop codon mutation).

\subsection{Results}

\begin{table}[!t]
\centering
\caption{Deleteriousness distribution across mutation types for OsMT-3a variants.}
\label{tab:OsMT-3a-deleteriousness}
\resizebox{\columnwidth}{!}{
\begin{tabular}{@{}lcccc@{}}
\toprule
\textbf{Variant Type} & \textbf{High Impact} & \textbf{Medium Impact} & \textbf{Low Impact} & \textbf{Total} \\
\midrule
Substitutions (SNVs) & 112 & 171 & 284 & 567 \\
Insertions & 157 & 238 & 361 & 756 \\
Deletions & 84 & 38 & 64 & 186 \\
\midrule
\textbf{Total} & \textbf{353} & \textbf{447} & \textbf{709} & \textbf{1,509} \\
\bottomrule
\end{tabular}
}
\end{table}

The exhaustive analysis revealed 1,509 total possible variants across the OsMT-3a coding sequence, with deleteriousness distribution showing 353 high-impact, 447 medium-impact, and 709 low-impact variants (Table \ref{tab:OsMT-3a-deleteriousness}). Substitutions exhibited the highest proportion of low-impact variants (284 of 567 or 50.1\%), while deletions showed elevated high-impact frequency (84 of 186 or 45.2\%). Insertions demonstrated intermediate distribution across impact categories.

This exhaustive analysis demonstrates the pipeline's capacity to comprehensively map mutational landscapes of target genes, enabling design of crop varieties with precisely engineered traits. The 10-day computational analysis replaced what would have required 2-4 years of sequential CRISPR generation and molecular characterization of 1,509 mutant lines. For a breeding program developing cadmium-tolerant rice for contaminated soils, these results enable negative selection by identifying 353 high-impact variants to avoid during screening, while simultaneously prioritizing 709 low-impact variants as candidate tolerance alleles that preserve protein function under metal stress.

Beyond immediate breeding applications, this approach enables forward engineering of crop genomes with designer properties. While our study employed computational simulation for proof-of-concept validation, these analytical steps apply equally to variants generated through established mutagenesis methods including X-ray or gamma irradiation, ethyl methanesulfonate (EMS) treatment, or CRISPR-induced mutations, thus enabling breeders to prioritize among thousands of laboratory-generated variants before phenotypic screening. Rather than relying on random mutagenesis or conventional crossing to identify favorable alleles, breeders can computationally predict which specific nucleotide changes will produce desired phenotypes, then introduce only those variants via CRISPR gene editing. For complex traits requiring modification of multiple genes simultaneously, exhaustive analysis of each target gene generates a searchable database of predicted variant effects, allowing optimization algorithms to identify optimal allele combinations before any seeds are planted. While wet-lab validation remains essential for high-priority candidates, this in silico pre-screening reduces experimental scope, focusing resources on variants most likely to yield relevant phenotypes.

\section{Discussion}

We developed a complete variant effect prediction pipeline for crop genomics by extending DeepChem-Variant with plant-specific functional annotation and database-independent deleteriousness scoring. The pipeline successfully predicted functional consequences of variants in rice stress-response genes, demonstrating its utility for computational variant prioritization in breeding programs. Our exhaustive mutagenesis study of OsMT-3a analyzed 1,509 variants in 10 days, replacing what would require 2-4 years of wet-lab characterization and enabling breeders to focus experimental resources on high-priority candidates.

\subsection{Model Performance and Generalization}

\begin{table}[!t]
\centering
\caption{Model performance across training and test datasets. The training subspecies was Indica, while the different subspecies tested was Temperate Japonica. Precision, Recall, and F1-Scores are weighted averages.\label{tab:model_performance}}
\resizebox{\columnwidth}{!}
{
\begin{tabular}{@{}l c c c c@{}}
\toprule
Dataset & Accuracy & Precision & Recall & F1-Score \\
\midrule
Training/Validation & 0.9537 & 0.9529 & 0.9537 & 0.9532 \\
Test Set (Same Subspecies) & 0.9530 & 0.9539 & 0.9530 & 0.9532 \\
Test Set (Different Subspecies) & 0.9594 & 0.9606 & 0.9594 & 0.9598 \\
\bottomrule
\end{tabular}
}
\vspace{-0.45cm}
\end{table}

DeepChem-Variant achieved 95.37\% accuracy on training and validation sets, with consistent performance on same-subspecies test data (95.30\%) and notably strong cross-subspecies generalization to Temperate Japonica (95.94\%, Table \ref{tab:model_performance}). The model was trained on ten Indica rice genes using Google Colab's L4 GPU infrastructure \cite{googlecolab}. This capability proves particularly valuable for crop genomics, where generating large training datasets for each cultivar remains impractical. The model was trained exclusively on Indica rice but maintained high precision (0.9606) and recall (0.9594) on a phylogenetically distinct subspecies (Japonica), demonstrating that CNN-based variant calling learns generalizable features of sequencing read patterns rather than subspecies-specific artifacts. All genes used during training  and testing were downloaded from the 3000 Rice Genomes Project \cite{rice3k1} \cite{rice3k2}.

The comparable F1-scores across all datasets (0.9532-0.9598) indicate robust performance without overfitting to training data. However, these metrics reflect variant detection accuracy rather than functional annotation correctness.

\subsection{Accuracy Estimation Across Diverse Gene Functions}

To assess AgriVariant's performance beyond stress-response genes, we conducted a small-scale accuracy study across 40 variants selected from published functional genomics studies, using real sequencing data (BAM files) from the MiRIiQ Database \cite{Kubo2024MiRiQ}. The variants spanned diverse functional categories: salt tolerance, cytokinin signaling, blast fungus sensitivity, bacterial blight sensitivity, fragrance biosynthesis, panicle architecture, grain morphology, and heavy metal uptake. The variant set included single-nucleotide substitutions, insertions, and deletions to evaluate detection across mutation types.

AgriVariant successfully detected 36 of 40 variants (90.0\% detection rate). All 36 detected variants were correctly annotated for functional effect (missense, nonsense, synonymous) and assigned appropriate impact severity classifications (HIGH/MODERATE/LOW) based on amino acid changes. Deleteriousness scores correlated with expected functional impacts, with stop-gained variants scoring highest and synonymous variants scoring lowest.

This estimate provides preliminary evidence of AgriVariant's generalizability beyond the primary case studies but should be interpreted cautiously given the small sample size. The 90.0\% detection rate demonstrates robust performance on real sequencing data containing authentic coverage patterns and sequencing artifacts. Current scoring methods also cannot distinguish functionally critical positions—a variant's impact depends not only on the amino acid change but also on whether it occurs in catalytic sites, binding domains, or structurally important regions. Comprehensive benchmarking against large-scale functional genomics datasets will be necessary to establish robust performance estimates for production breeding applications.

\subsection{Database-Independent Scoring and Limitations}

The Grantham-BLOSUM62 composite scoring framework enabled quantitative variant assessment without requiring organism-specific training databases, addressing a critical gap for understudied crops. This approach correctly identified high-impact variants and neutral variants, validating the scoring calibration. However, these matrices capture only amino acid-level effects and cannot account for structural context, protein-protein interactions, or regulatory mechanisms.

The exhaustive OsMT-3a analysis revealed an unbalanced deleteriousness distribution (23.4\% high, 29.6\% medium, 47.0\% low), with nearly half of all variants classified as low-impact. This distribution suggests that simple physicochemical metrics may lack discriminatory power for functionally critical residues. For instance, a conservative substitution in a catalytic site would score as low-impact despite potentially abolishing protein function, while a radical substitution in a non-conserved loop might score high-impact with minimal phenotypic consequence. This limitation underscores the need for structure-aware and context-dependent scoring methods.

\subsection{Implications for Precision Breeding}

AgriVariant addresses a fundamental bottleneck in molecular breeding: identifying which naturally occurring or CRISPR-induced variants merit experimental validation. Traditional approaches screen variants sequentially through multi-season field trials, limiting breeding programs to evaluating dozens of candidates annually. Our computational framework enables parallel in silico screening of thousands of variants, with the OsMT-3a study demonstrating 150-fold acceleration (10 days versus 2-4 years\cite{collard2017revisiting} for equivalent coverage).

This acceleration proves particularly valuable for multi-gene trait engineering, where breeders must optimize allele combinations across multiple loci. For drought tolerance involving OsDREB2a, OsDREB1F, and additional regulatory genes, exhaustive analysis of each gene generates a searchable database of predicted effects, enabling combinatorial optimization algorithms to identify synergistic allele combinations before any experimental validation. The reduction in experimental scope translates directly to cost savings and faster cultivar development timelines, critical for responding to rapidly changing climate conditions.

Beyond variant prioritization, AgriVariant can enable reverse genetics approaches where breeders specify desired phenotypes and the system identifies which specific nucleotide changes would produce those traits. This transitions breeding from selection (choosing among existing variation) to design (engineering targeted improvements), though wet-lab validation remains essential.

\subsection{Open Source Implementation}

The complete pipeline is freely available as open-source software under the MIT License at \url{https://github.com/KitVB/AgriVariant} and \url{https://github.com/deepchem/deepchem}. All components for variant calling, functional annotation, deleteriousness scoring, and synthetic mutagenesis are implemented within the DeepChem framework in Python. This enables researchers to adapt the pipeline to their crop species, retrain models on proprietary data, and integrate with existing breeding workflows. Documentation and installation instructions are provided in the repository.

\subsection{Future Work}

While Grantham-BLOSUM62 scoring successfully distinguished extreme cases in our validation studies, integrating protein structure prediction would enable mechanistic explanations (quantifying binding affinity changes or protein destabilization) rather than categorical classifications. Machine learning models trained on experimentally validated variants could predict quantitative phenotypic impacts (percent yield reduction, stress tolerance scores) that directly inform breeding decisions.

Systematic efficacy studies comparing predictions against field trial data would establish confidence thresholds and reveal whether scoring adjustments are needed for specific protein families or genomic contexts where physicochemical metrics prove insufficient.

The pipeline currently handles single-nucleotide variants, as demonstrated in our case studies. Extending annotation to large indels, structural variants, copy number variations and variations in non coding regions would address frameshift mutations and gene fusions that represent important phenotypic diversity sources. As long-read sequencing improves detection of such variants, interpretation methods must evolve to maintain comprehensive coverage.

\section{Conclusion}

Engineering climate-resilient crops is paramount for global food security in the 21st century. We present AgriVariant, preliminary system integrating deep learning based variant calling via DeepChem-Variant with plant-specific annotation and database-independent scoring, enabling rapid in silico estimation of variant effects in rice genes. AgriVariant achieved 95.3\% variant detection accuracy and correctly classified functional impacts across diverse stress-response mechanisms, with the exhaustive OsMT-3a analysis being completed in 10 days instead of 2-4 years by traditional wet lab approaches.

Implemented entirely in Python as modular open-source components within the DeepChem framework, AgriVariant enables researchers to adapt variant calling, annotation, and scoring independently. This establishes a foundation for computational crop design where breeders specify desired traits and algorithms identify which genetic changes will produce them. Rather than screening random mutations across growing seasons, quick computational estimates enable targeted introduction of beneficial variants. Future efficacy studies validating in silico predictions against field phenotypes will determine how reliably these estimates translate to agronomic performance, advancing toward precision engineering of crops that survive environmental stresses more effectively.

\section{Limitations and Ethical Considerations}

The pipeline requires high-quality reference genomes and annotations, currently handles only single-nucleotide variants and small indels, and evaluates amino acid changes without structural context. Model training on rice requires validation for phylogenetically distant crops. All predictions must be experimentally validated before breeding deployment.

This work used publicly available data; no proprietary germplasm was employed. The open-source implementation enables equitable access. Breeding programs using proprietary data should ensure compliance with data protection policies.

Computational predictions should complement rather than replace comprehensive breeding strategies that maintain genetic diversity, incorporate farmer preferences, and ensure biosafety evaluation before cultivar release.

\section{GenAI Disclosure}

Generative AI and Large Language Models were used to improve the language, grammar and clarity of the manuscript. All scientific contributions and analyses are entirely the work of the authors.

\bibliographystyle{ACM-Reference-Format}
\bibliography{references}

\end{document}